\documentclass[12pt, referee]{aastex}
\usepackage[dvips]{color}
\newcommand{\bff}{}
\newcommand{\re}{}
\def\ltsima{$\;\buildrel < \over \sim \;$}
\def\simlt{\lower.5ex \hbox{\ltsima}}
\def\gtsima{$\;\buildrel > \over \sim \;$}
\def\simgt{\lower.5ex \hbox{\gtsima}}

\shorttitle{Interstellar chloronium - II.}
\shortauthors{Neufeld et al.}

\begin{document}

\title{{\it Herschel}$^*$ observations of interstellar chloronium.  II -- Detections toward G29.96-0.02, W49N, W51, and W3(OH), and determinations of the ortho-to-para and $^{35}$Cl/$^{37}$Cl isotopic ratios }
\author{David A.~Neufeld\altaffilmark{1}, John~H.~Black\altaffilmark{2}, Maryvonne~Gerin\altaffilmark{3,4}, Javier~R.~Goicoechea\altaffilmark{5}, Paul~F.~Goldsmith\altaffilmark{6}, Cecile~Gry\altaffilmark{7}, Harshal Gupta\altaffilmark{8}, Eric~Herbst\altaffilmark{9}, Nick~Indriolo\altaffilmark{10}, Dariusz~Lis\altaffilmark{3,8}, Karl~M.~Menten\altaffilmark{11}, Raquel~Monje\altaffilmark{8}, Bhaswati~Mookerjea\altaffilmark{12}, Carina~Persson\altaffilmark{2}, Paule~Sonnentrucker\altaffilmark{13}, and Mark~G.~Wolfire\altaffilmark{14}}

\altaffiltext{*}{Herschel is an ESA space observatory with science instruments provided
by European-led Principal Investigator consortia and with important
participation from NASA}
\altaffiltext{1}{\small Department of Physics \& Astronomy, Johns Hopkins University,
3400~N.~Charles~St., Baltimore, MD 21218, USA}
\altaffiltext{2}{\small Dept.\ of Earth and Space Sciences, Chalmers University of Technology, Onsala, Sweden}
\altaffiltext{3}{\small LERMA, Observatoire de Paris, PSL Research University, CNRS, Sorbonne Universit\'es, UPMC Univ. Paris 06, F-75014, Paris, France}
\altaffiltext{4}{\small \'Ecole normale sup\'erieure, F-75005, Paris, France}
\altaffiltext{5}{\small Grupo de Astrof\'isica Molecular. Instituto de Ciencia de Materiales de Madrid (ICMM, CSIC). Sor Juana Ines de la Cruz 3, 28049 Cantoblanco, Madrid, Spain}
\altaffiltext{6}{\small Jet Propulsion Laboratory, California Institute of Technology, 4800 Oak Grove Drive, Pasadena, CA 91109-8099, USA}
\altaffiltext{7}{\small Aix Marseille Universit\'é, CNRS, LAM (Laboratoire d'Astrophysique de Marseille) UMR 7326, 13388, Marseille, France}
\altaffiltext{8}{\small California Institute of Technology, Pasadena, CA 91125, USA}
\altaffiltext{9}{\small Departments of Chemistry, Astronomy, and Physics, University of Virginia, Charlottesville, VA 22904, USA}
\altaffiltext{10}{\small Department of Astronomy, University of Michigan, Ann Arbor, MI 48109, USA}
\altaffiltext{11}{\small Max-Planck-Institut f\"ur Radioastronomie, Auf dem H\"ugel 69, 53121 Bonn, Germany}
\altaffiltext{12}{\small Tata Institute of Fundamental Research, Homi Bhabha Road, Mumbai 400005, India}
\altaffiltext{13}{\small Space Telescope Science Institute, 3700 San Martin Drive, Baltimore, MD 21218, USA}
\altaffiltext{14}{\small Department of Astronomy, University of Maryland, College Park, MD 20742, USA}

\begin{abstract}

{We report additional detections of the chloronium molecular ion, H$_2$Cl$^+$, toward four bright submillimeter continuum sources: G29.96, W49N, W51, and W3(OH).  With the use of the HIFI instrument on the {\it Herschel Space Observatory}, we observed the $2_{12}-1_{01}$ transition of ortho-H$_2^{35}$Cl$^+$ at 781.627~GHz 
in absorption toward all four sources.  Much of the detected absorption arises in diffuse foreground clouds that are unassociated with the background continuum sources and in which our best estimates of the $N({\rm H_2Cl^+})/N({\rm H})$ ratio lie in the range \re{(0.9 -- 4.8)} $\times 10^{-9}$.  \re{These} chloronium abundances relative to atomic hydrogen can exceed the predictions of current astrochemical models by up to a factor of 5.
Toward W49N, we have also detected the $2_{12}-1_{01}$ transition of ortho-H$_2^{37}$Cl$^+$ at 780.053~GHz and the $1_{11}-0_{00}$ transition of para-H$_2^{35}$Cl$^+$ at 485.418~GHz.  These observations imply  H$_2^{35}$Cl$^+$/H$_2^{37}$Cl$^+$ \re{column density} ratios that are consistent with the solar system $^{35}$Cl/$^{37}$Cl isotopic ratio of 3.1, and chloronium ortho-to-para ratios consistent with 3, the ratio of spin statistical weights.}

\end{abstract}


\section{Introduction}

The chloronium molecular ion, H$_2$Cl$^+$, is one of several new interstellar molecules  discovered using the HIFI instrument (de Graauw et al.\ 2010) on board the {\it Herschel Space Observatory} (Pilbratt et al.\ 2010).
\re{Chloronium} was first detected by Lis et al.\ (2010) -- who used the {\em Herschel}/HIFI to perform absorption line spectroscopy of the $1_{11}-0_{00}$ 485~GHz transition of para-chloronium toward the bright submillimeter continuum sources NGC 6634I and Sgr B2 (S).  \re{Toward NGC 6634I, the $2_{12}-1_{01}$ 782~GHz transition of ortho-chloronium was also detected, leading to an inferred ortho-to-para ratio of 3.2, close to the ratio of statistical weights for the two spin states.}  Interstellar H$_2$Cl$^+$ was subsequently detected by Neufeld et al.\ 2012 (Paper I) in absorption toward W31C and Sgr A ($+50\,\rm km\,s^{-1}$ cloud), and in emission from the Orion Bar photodissociation region and Orion South molecular condensation.  The chloronium abundances inferred from the absorption line detections in Paper I, amounting to as much as 12$\%$ of the gas-phase chlorine budget, were surprisingly high relative to predictions \re{obtained using the Meudon PDR model (Le Petit et al.\ 2006) with a chemical network for Cl-bearing molecules similar to that described by Neufeld \& Wolfire (2009, hereafter NW09).}  These {\em Herschel} observations were followed by ground-based (IRAM 30-m and CSO) detections of the 189~GHz $1_{10}-1_{01}$ transition of ortho-H$_2$Cl$^+$ toward W31C and W49N (Gerin et al.\ 2013), which -- \re{combined} with {\it Herschel}/HIFI detections of the $1_{11}-0_{00}$ transition of para-H$_2$Cl$^+$ -- provided 
 \re{additional} 
 estimates of the H$_2$Cl$^+$ ortho-to-para ratio.  These estimates were consistent with the high-temperature LTE value of 3, but had very large error bars.  Most recently, extragalactic chloronium has been detected in the $z = 0.89$ absorber in front of the lensed blazar PKS 1830--211 (Muller et al.\ 2014), again through observations of the  $1_{11}-0_{00}$ transition, redshifted now to $\sim 257$~GHz.
 
{\bff The chemistry of interstellar chloronium has been discussed extensively in NW09 and Paper I and only a brief summary is needed here.  With an ionization potential (12.97 eV) slightly below that of atomic hydrogen, chlorine is singly-ionized in diffuse atomic clouds.  In regions where the H$_2$ fraction becomes significant, Cl$^+$ reacts exothermically with H$_2$ to form HCl$^+$, a molecule that has now been detected (de Luca et al.\ 2012) in the diffuse interstellar medium by {\it Herschel}/HIFI.  A further exothermic reaction with H$_2$ leads to H$_2$Cl$^+$, which ultimately undergoes dissociative recombination to form HCl or Cl.  Thus the overall pathway leading to chloronium may be written: $\rm Cl\,(h\nu,e)\,Cl^+(H_2,H)\,HCl^+(H_2,H)\,H_2Cl^+$. The chloronium abundance is limited by the dissociative recombination of HCl$^+$ and H$_2$Cl$^+$, and by the relatively modest flux of UV photons of sufficient energy to initiate the reaction sequence by ionizing Cl.  Detailed predictions for the depth dependence of the chloronium abundance have been presented by NW09, but the predicted column densities were up to an order of magnitude lower than the observed values reported in Paper I.  Although several possible explanations of this discrepancy were investigated in Paper I, none proved entirely satisfactory and thus the high chloronium abundances measured toward Sgr A and particularly W31C remain puzzling.  Like other triatomic hydrides such as H$_2$O, H$_2$O$^+$ and H$_2$S, chloronium possesses two symmetry states -- ortho-H$_2$Cl$^+$ with total H nuclear spin of 1, and para-H$_2$Cl$^+$ with total spin 0 -- each associated with distinct sets of rotational states of even (ortho-H$_2$Cl$^+$) and odd (para-H$_2$Cl$^+$) spatial symmetry.}

Among all the elements lighter than nickel, chlorine possesses the minor isotope ($^{37}$Cl) that is most abundant relative to its principal isotope ($^{35}$Cl), with a solar system $^{35}$Cl/$^{37}$Cl ratio of only 3.1 (Lodders et al.\ 2003).\footnote{\footnotesize This fact, resulting in an atomic weight of 35.45 for terrestrial chlorine that was unexplained until the discovery of isotopes by \re{Aston} in 1919, was of historical importance in Turner's (1833) refutation of Prout's Law (which had asserted that the atomic weights of the elements were integral multiples of that of hydrogen).}  Thus, the  H$_2^{37}$Cl$^+$ isotopologue is often detected together with H$_2^{35}$Cl$^+$, \re{the frequencies for both isotopologues falling easily within the simultaneous bandwidth of the instrument}; indeed, the H$_2^{37}$Cl$^+$ $1_{11}-0_{00}$ transition was detected in the NGC 6334I, W31C, Sgr B2 and PKS 1830--211 observations discussed above.  In the absence of isotope-selective fractionation in the interstellar gas, the H$_2^{35}$Cl$^+$/H$_2^{37}$Cl$^+$ ratio serves as a valuable probe of the $\rm ^{35}Cl/^{37}Cl$ isotopic ratio outside the solar system; in all cases reported to date, including the $z = 0.89$ absorber in front of blazar PKS 1830--211, the H$_2^{35}$Cl$^+$/H$_2^{37}$Cl$^+$ ratio is consistent, within the observational uncertainties, with the solar system $\rm ^{35}Cl/^{37}Cl$ isotopic ratio.

In this paper, we report new {\it Herschel}/HIFI observations of interstellar chloronium.  We have targeted the $2_{12}-1_{01}$ transition of ortho-H$_2$Cl$^+$ near 781~GHz, which presents a larger optical depth than either of the 485~GHz or 189~GHz transitions observed previously.  In \S 2 below, we discuss the observations and data reduction, which have led to unequivocal detections of ortho-H$_2^{35}$Cl$^+$ absorption in foreground diffuse clouds lying along the sight-lines to the bright submillimeter continuum sources G29.96, W49N, W51, and W3(OH), together with unequivocal detections of ortho-H$_2^{37}$Cl$^+$ and para-H$_2^{35}$Cl$^+$ toward W49N.  The resultant spectra, and the inferred chloronium column densities and abundances, are presented in \S 3, together with the ortho-to-para ratios and  H$_2^{35}$Cl$^+$/H$_2^{37}$Cl$^+$ ratios derived for clouds along the sight-line to W49N.  A discussion of these results follows in \S 4. 
   
\section{Observations and data reduction}

All the data presented here were obtained as part of the program OT1\_dneufeld\_1 on 2013 March 26 -- 28 in the final weeks of the {\it Herschel} mission.  The $2_{12}-1_{01}$ transitions of ortho-H$_2^{35}$Cl$^+$ and H$_2^{37}$Cl$^+$ were observed simultaneously in the lower sideband of Band 4a of HIFI, and the $1_{11}-0_{00}$ transitions of para-H$_2^{35}$Cl$^+$ and H$_2^{37}$Cl$^+$ were observed in the lower sideband of Band 1a.  These ortho- and para-chloronium transitions were targeted in observations of total duration 2545~s and 3920~s respectively toward each source, \re{which provided spectra with r.m.s. noise levels of $\sim 10$~mK and $\sim 4$~mK, respectively.}  The observing time was divided among three separate {\bff local oscillator} (LO) settings; because HIFI employs double sideband receivers, the use of multiple LO settings allows one to determine whether any given feature lies in the upper or lower sideband of the receiver.  Table 1 lists, for each target, the coordinates, distance, {\it Herschel} observation identifiers (OBSID), observed continuum brightness temperatures, and source systemic velocity relative to the local standard of rest (LSR).  Table 2 lists the rest frequencies for the $2_{12}-1_{01}$ and $1_{11}-0_{00}$ transitions of each isotopologue, using data downloaded from the Cologne Database for Molecular Spectroscopy (CDMS; M\"uller et al.\ 2005), which are based on laboratory spectroscopy performed by Araki et al.\ (2001) and a theoretical estimate of the dipole moment by M\"uller (2008).  Because $^{35}$Cl and $^{37}$Cl both possess a nuclear spin of 3/2, the $2_{12}-1_{01}$ transition is split into eight hyperfine components and the $1_{11}-0_{00}$ transition into three.

The data were reduced using methods similar to those discussed in Paper I.  
Because of the presence of strong interloper emission lines and/or the small signal-to-noise ratios obtained in some of the observed spectra, only six of the 16 possible combinations of target line and background source proved to be of value in providing quantitative results: the ortho-H$_2^{35}$Cl$^+$ spectra observed toward all four background sources, and the ortho-H$_2^{37}$Cl$^+$ and para-H$_2^{35}$Cl$^+$ spectra obtained toward W49N.  In Figures 1 through 3, we show the six spectra listed above, with the three LO settings shown separately in red, green and blue.  {\bff Here, the spectra obtained for the two orthogonal polarizations were found to be very similar and were averaged to reduce the noise level, yielding signal-to-noise ratios on the continuum in the range $\sim 100 - 300$.}
Alternate panels show the frequencies computed for either the signal (black frames) or image sideband (brown frames) in the rest frame of the source.  Magenta dashed lines indicate identified emission lines in the relevant sideband.  

The different spectra obtained at three separate LO settings and in two orthogonal polarizations were averaged with a weighting inversely proportional to the square of the r.m.s. noise, except \re{for spectral regions} where a contaminating emission feature from the image sideband caused interference with the H$_2$Cl$^+$ absorption feature; \re{such regions were} excluded from the average.  In the para-H$_2^{35}$Cl$^+$ spectra observed toward W49N, a ripple is clearly apparent in the data and is a known artifact of the instrument (Roelfsema et 
al.\ 2012).  It was removed by using the task {\it fitHiFiFringe} within the Herschel Interactive Processing Environment version 12.1.0 (HIPE; Ott 2010) to subtract standing waves at 92 and 98 \re{MHz}.  Finally, the averaged absorption spectra were converted to a transmission fraction under the assumption that the sideband gain is unity; the transmission is then given by  $2 T_A/T_A({\rm cont}) - 1$, where $T_A$ is the double-sideband antenna temperature and $T_A({\rm cont})$ is the continuum temperature.

\section{Results}

In Figure 4, we show (black histograms) the transmission fractions for ortho-H$_2^{35}$Cl$^+$ observed toward all four background sources, and for ortho-H$_2^{37}$Cl$^+$ and para-H$_2^{35}$Cl$^+$ observed toward W49N.  All six panels indicate widespread absorption by foreground material unassociated with the background continuum source.  We used a multigaussian fit to the absorbing column density to deconvolve the hyperfine structure, with the fit to the data shown in 
red and the hyperfine-deconvolved fit in blue.  \re{The latter is the spectrum that would have resulted in the absence of hyperfine splitting, and shows the same frequency-integrated optical depth as the original data.} The emission feature appearing at LSR velocity $\sim 40\,\rm km\,s^{-1}$ in the G29.96--0.02 ortho-H$_2^{35}$Cl$^+$ spectrum and at LSR velocity $\sim 0\,\rm km\,s^{-1}$ in the W51 spectrum is a blend of dimethyl ether transitions lying in the same sideband as the ortho-H$_2^{35}$Cl$^+$ transition; spectral regions within $10\,\rm km\,s^{-1}$ of that feature were excluded from the hyperfine-deconvolution procedure.

In Table 3, we present the column densities derived for ortho-H$_2^{35}$Cl$^+$ within those velocity intervals adopted by Indriolo et al.\ (2015) for each source. These column densities were derived 
\re{for an assumed excitation temperature of 2.73~K, the temperature of the cosmic microwave background,}
an approximation that is expected to be excellent for foreground diffuse clouds where the density and radiation field are low, but one that \re{is likely} invalid for material associated with the background source itself.  \re{For an excitation temperature of 2.73~K, the ratio of the ortho-$\rm H_2Cl^+$ column density to the total, velocity-integrated optical depth for all hyperfine components in the $2_{12}-1_{01}$ transition is $\rm 4.65 \times 10^{12}\,cm^{-2}\, per \, km\,s^{-1}$, while the ratio of the para-$\rm H_2Cl^+$ column density to the total, velocity-integrated optical depth for all hyperfine components in the $1_{11}-0_{00}$ transition is $\rm 2.24 \times 10^{12}\,cm^{-2} \, per \, km\,s^{-1}$.}

Table 3 also lists the HI column densities given by Indriolo et al.\ (2015) for each velocity interval, based upon recent 21 cm data obtained by Winkel et al.\ (2015), and the corresponding $N({\rm H_2Cl^+})/N({\rm H})$ column density ratios, for an assumed ortho-to-para ratio of 3 and an assumed H$_2^{35}$Cl$^+$/H$_2^{37}$Cl$^+$ ratio of 3.1.   These results are presented graphically in Figure 5.  Here, dashed horizontal bars indicate velocity intervals close the source systemic velocity, where the column density determinations may be unreliable, and vertical error bars indicate uncertainties.  The latter are highly asymmetric because they include the possibility that weak unidentified emission lines in the relevant spectral window result in an underestimate of the inferred H$_2$Cl$^+$ optical depth. \re{Based upon an examination of the entire spectra shown in Figure 1 -- 3, we conservatively adopt a maximum antenna temperature of 50 mK and linewidth of 5$\,\rm km\,s^{-1}$ for any such emission lines in ortho-chloronium spectra; for the para-chloronium spectrum obtained toward W49N, we adopt instead a maximum antenna temperature of 25 mK.}

In Table 4, we compare the column densities derived for ortho-H$_2^{35}$Cl$^+$, ortho-H$_2^{37}$Cl$^+$, and para-H$_2^{35}$Cl$^+$ toward W49N.  The corresponding ortho-to-para ratios and H$_2^{35}$Cl$^+$/H$_2^{37}$Cl$^+$ ratios are also tabulated for the velocity intervals adopted by Indriolo et al.\ (2015).  In Figure 6, these ratios are plotted in velocity bins of width $5\,\rm km\,s^{-1}$, with vertical error bars again indicating the uncertainties that could result from unidentified emission lines in the spectral window (with a width of 5$\,\rm km\,s^{-1}$ and an assumed maximum antenna temperature of 50 mK near 781~GHz and 25 mK near 485~GHz).  Values close the systemic velocity of the source  ($8\,\rm km\,s^{-1}$) \re{are} unreliable.

\section{Discussion}

\subsection{Chloronium abundances relative to atomic hydrogen}

Our best estimates of the $N({\rm H_2Cl^+})/N({\rm H})$ column density ratios, listed in Table 3 and plotted in Figure 5, lie in the range $\re{(0.87 - 4.84)} \times 10^{-9}$.  This range excludes velocity intervals close to the systemic velocity of the background source, where the $N({\rm H_2Cl^+})$ estimates are unreliable.
Theoretical predictions, presented previously in Paper I and based upon the Meudon PDR code (Le Petit et al.\ 2006) and the chemical network of Neufeld \& Wolfire (2009) for the chemistry of halogen-bearing molecules in diffuse clouds can predict $N({\rm H_2Cl^+})/N({\rm H})$ ratios as high as $1 \times 10^{-9}$, given a gas-phase elemental abundance of  $1 \times 10^{-7}$ for chlorine (Moomey et al.\ 2012).  Thus, some of the observed  $N({\rm H_2Cl^+})/N({\rm H})$ ratios exceed the predictions by a factor of at least 5, a discrepancy noted previously toward other sources and discussed in Paper I.  Moreover, if weak, unidentified emission lines are present within the spectral region where $N({\rm H_2Cl^+})$ absorption is observed, the actual ${\rm H_2Cl^+}$ abundances might significantly exceed our best estimates, as indicated by the asymmetric error bars in Figure 5.  

None of our best estimates for the $N({\rm H_2Cl^+})/N({\rm H})$ ratio is as large as the value of $12 \times 10^{-9}$ reported in Paper I for the sight-line to W31C.  However, that large value may have been based upon an underestimate of the HI column density toward W31C, which had been estimated from 21 cm line spectra presented by Fish et al.\ (2003).  More recent HI 21 cm observations, carried out by Winkel et al.\ (2015) using the JVLA, have suggested that the 21 cm absorption may in fact be saturated along the sight-line to W31C, providing only lower limits on the HI density (Indriolo et al.\ 2015) and therefore upper limits on the $N({\rm H_2Cl^+})/N({\rm H})$ ratio. 

If we now disregard the large $N({\rm H_2Cl^+})/N({\rm H})$ ratio reported for W31C, the maximum discrepancy between theory and observation is reduced from the factor $\sim 10$ discussed in Paper I to a factor $\sim 5$.  Moreover, since the publication of Paper I, Novotn{\'y}  et al.\ (2013) have reported a laboratory measurement of the rate coefficient for dissociative recombination (DR) of HCl$^+$, a key intermediary in the primary production pathway for chloronium: $\rm Cl\,(h\nu,e)\,Cl^+(H_2,H)\,HCl^+(H_2,H)\,H_2Cl^+$.  The measured value is smaller than that adopted by NW09 by \re{a factor of} 1.5 at 10~K and 3 at 300~K.  Since the DR of HCl$^+$ limits the availability of HCl$^+$ to form $\rm H_2Cl^+$, a downward revision in the adopted rate coefficient would increase the 
predicted $\rm H_2Cl^+$ abundances; however, the dependence is a relatively weak one, in that a factor 10 reduction in that rate coefficient was found to increase the $\rm H_2Cl^+$ abundance by only a factor of 1.3 (Paper I).  Definitive measurements of the DR rates for HCl$^+$ under astrophysical conditions -- and also for $\rm H_2Cl^+$, which is primarily destroyed by DR -- will have to await future experiments at the Cryogenic Storage Ring (CSR) in Heidelberg.  This facility will permit the first storage ring measurements of {\it rotationally-cold} molecular ions, in contrast to previous measurements in which the recombining ions had a rotational temperature $\sim 300$~K.  The case of rotationally-cold ions is most appropriate to hydride molecular ions in low-density diffuse clouds, where most ions are in the ground rotational states; depending upon the exact details specific to each DR process, future CSR measurements could yield rate coefficients that are either smaller {\it or} larger than those obtained with rotationally-warm ions.

The astrochemical models presented in Paper I predict that the maximum $N({\rm H_2Cl^+})/N({\rm H})$ ratios are attained in diffuse clouds of total visual extinction ($\sim 0.3$~mag) in which the gas is primarily atomic, a prediction that is supported by the observed distribution of H$_2$Cl$^+$ in velocity space.  In particular, the ${\rm H_2Cl^+}$ absorption line profiles plotted in Figure 1 are more similar to those of OH$^+$ and H$_2$O$^+$ (Indriolo et al.\ 2015) -- two species believed to originate in gas of small molecular fraction -- than they are to those of H$_2$O, HF (Sonnentrucker et al.\ 2015) or CS (Neufeld et al.\ 2015). \re{The} latter three molecules, believed to be most abundant in gas with a high molecular fraction, tend to show narrow absorption features that are entirely absent in the OH$^+$, H$_2$O$^+$ and ${\rm H_2Cl^+}$ spectra.

\subsection{Chloronium isotopologue ratio}

Because isotope-selective fractionation is expected to be entirely negligible, the 
$N({\rm H_2^{35}Cl^+)}$/$N({\rm H_2^{37}Cl^+)}$ ratio provides a valuable measure of the $^{35}$Cl/$^{37}$Cl isotopic ratio within the interstellar gas.  This probe has been exploited previously by Lis et al.\ (2010) and in Paper I, with observations of para-chloronium absorption implying best-fit $^{35}$Cl/$^{37}$Cl ratios in the range $\sim 2 - 4$ within foreground clouds along the sight-lines to NGC 6334I, W31C and Sgr A ($+50\,\rm km\,s^{-1}$ cloud); these values bracket the solar system $^{35}$Cl/$^{37}$Cl ratio of 3.1 (Lodders 2003).  
Similar results have been obtained from observations of HCl emission toward W3 A (Cernicharo et al.\ 2010) and several other Galactic sources (Peng et al.\ 2010).   In the one extragalactic determination of a $\re{N}({\rm H_2^{35}Cl^+)}$/$\re{N}({\rm H_2^{37}Cl^+)}$ ratio obtained to date, Muller et al. (2014) reported a value of $3.1_{-0.2}^{+0.3}$ for the $z=0.89$ absorber in front of PKS 1830--211, based on absorption-line observations of para-chloronium.  In the present study, our observations of the stronger ortho-chloronium transitions towards W49N permit more accurate determinations of the $^{35}$Cl/$^{37}$Cl ratio than were possible in Paper I.  As shown in Figure 6 (middle panel), the $N({\rm H_2^{35}Cl^+)}$/$N({\rm H_2^{37}Cl^+)}$ ratios determined in $5\,\rm km\,s^{-1}$ velocity bins are all consistent with the solar system $^{35}$Cl/$^{37}$Cl isotopic ratio to within the uncertainties.  The average for the entire sight-line is $3.50^{+0.21}_{-0.62}$ (Table 4).

The solar system $^{35}$Cl/$^{37}$Cl ratio is not entirely understood.  Citing Galactic chemical evolution models of Kobayashi, Karakas \& Umeda (2011; hereafter KKU11), Muller et al.\ (2014) noted that the isotopic ratios for most elements in the $z=0.89$ absorber in front of PKS 1830--211 were fitted well by a KKU11 model for stars in the solar neighborhood with [Fe/H] = --2.6 (i.e.\ stars with iron abundances, relative to H, a factor $10^{-2.6}$ times the solar value).  For that model, in which Type II supernovae were the source of heavy elements, the predicted $^{35}$Cl/$^{37}$Cl isotopic ratio was 3.0, very close to the solar system value of 3.1 and entirely consistent with the value of $3.1_{-0.2}^{+0.3}$ for the $z=0.89$ absorber in front of PKS 1830--211. However, the KKU11 model with [Fe/H] = 0.0 (i.e.\ with solar abundances) predicts a $^{35}$Cl/$^{37}$Cl isotopic ratio of only 1.9, a factor $\sim 1.6$ below the solar system value, and a value that is significantly discrepant with the results we have obtained here for the Galactic ISM.  This modest discrepancy notwithstanding, the KKU11 model does predict that the $^{35}$Cl/$^{37}$Cl will show a relatively weak dependence on [Fe/H], in sharp contrast to most of the other elemental isotopic ratios considered.  Thus, although the solar $^{35}$Cl/$^{37}$Cl ratio is evidently not predicted with perfect accuracy, the model does correctly suggest that $^{35}$Cl/$^{37}$Cl will not evolve strongly.

\subsection{Chloronium ortho-to-para ratio}

The ortho-to-para ratios (OPRs) plotted in Figure 6 (bottom panel) are all consistent with the ratio of statistical weights, 3, to within the estimated uncertainties.  This is the behavior expected in LTE at temperatures much larger than $\Delta E_{\rm op}/k = 20.2$~K, where $\Delta E_{\rm op}$ is the energy difference between the lowest rotational states of ortho- and para-chloronium ($1_{01}$ and $0_{00}$ respectively.)  Observations of some other polyatomic hydrides in diffuse molecular clouds reveal OPRs consistent with the ratio of statistical weights -- examples include H$_2$O (Flagey et al.\ 2013; Lis et al.\ 2013) and H$_2$O$^+$ (Schilke et al.\ 2013; Gerin et al.\ 2013) -- but this behavior is not universal.  In particular, $\rm H_3^+$ (Crabtree et al.\ 2011) and $\rm NH_3$ (e.g. Persson et al.\ 2012) both show OPRs with significant departures from the values expected in LTE, a behavior that is believed to reflect the OPR of H$_2$ molecules involved in the formation of these hydrides through nuclear-spin-conserving gas-phase reactions (e.g.\ Oka 2004, Faure et al.\ 2013; Le Gal et al.\ 2014).  

When chloronium is formed by the reaction $\rm HCl^+(H_2,H)H_2Cl^+$ -- or indeed when any triatomic hydride molecular ion (such as H$_2$O$^+$) is formed by an analogous hydrogen abstraction reaction -- the resultant ions are formed with an initial ortho-to-para ratio, OPR$_0$, that is determined by the nature of the intermediate complex; if the latter is long-lived, then the reaction occurs by the scrambling mechanism (e.g. Oka 2004, Herbst \re{2015}), in which all routes that lead to products are allowed.  In this case, one-half of all reactions of $\rm HCl^+$ with para-H$_2$ lead to ortho-$\rm H_2Cl^+$ (with the remainder forming para-$\rm H_2Cl^+$), and five-sixths of all reactions of $\rm HCl^+$ with ortho-H$_2$ lead to ortho-$\rm H_2Cl^+$.  Thus, $\rm H_2Cl^+$ is formed with an initial ortho-to-para ratio, $\rm OPR_0$, that is related to the OPR of H$_2$:
$$\rm OPR_0 (H_2Cl^+) = {5 \, OPR (H_2) + 3 \over OPR (H_2) + 3}. \eqno(1)$$
However, if the formation mechanism does not involve a long-lived intermediate complex, but occurs instead by a hopping mechanism in which a hydrogen atom breaks off from H$_2$ and attaches to  HCl$^+$ in a long-range interaction, then $\rm OPR_0(H_2Cl^+)$ is the ratio of statistical weights (i.e.\ 3), regardless of $\rm OPR(H_2)$ .  Detailed quantal  calculations will be needed to determine which case applies to H$_2$Cl$^+$.

As discussed by Gerin et al.\ (2013) and Herbst \re{(2015)}, the relationship between $\rm OPR_0(H_2Cl^+)$ and the observed OPR for H$_2$Cl$^+$ depends upon the relative rates of two processes: (1) destruction, which  occurs primarily via dissociative recombination at a rate that we assume to be identical for the two spin symmetries; and (2) thermalization via reactive collisions:
$$\rm ortho-H_2Cl^+ + H \,\, \re{\leftrightharpoons} \,\, para-H_2Cl^+ + H. \eqno(2)$$    The latter process tends to drive $\rm OPR(H_2Cl^+)$ to its LTE value, $\rm OPR_{LTE}(H_2Cl^+)$.  The solid black curve in Figure 7 shows the dependence of $\rm OPR_{LTE}(H_2Cl^+)$ upon the gas temperature, while the red curve shows the analogous quantity for molecular hydrogen, $\rm OPR_{LTE}(H_2).$  Dashed red and black curves show the functions 
$9\, \exp(-20.2 {\rm K}/T)$ and $9\, \exp(-170 {\rm K}/T)$ applying in the low temperature limit where only the lowest two rotational states are populated.

Given an H$_2$ OPR that is in LTE -- the behavior typically predicted in theoretical models for diffuse molecular clouds -- and if the formation of 
$\rm H_2Cl^+$ occurs via the scrambling mechanism, then the blue curve shows $\rm OPR_0 (H_2Cl^+)$, as computed from equation (1).  The observed $\rm OPR(H_2Cl^+)$ is therefore expected to lie between the blue and the black curves, depending upon the relative rates of destruction and thermalization.  Typically, astrochemical models (e.g. NW09) predict that chloronium is most abundant in gas that is primarily atomic and has a temperature $\sim 100$~K.

In steady-state, and with the assumptions given above, we find that  
$${\rm OPR(H_2Cl^+)} = {k_{\rm po} n({\rm H}) + k_{\rm dr} n_e {\rm OPR_0 (H_2Cl^+)} \over k_{\rm op} n({\rm H}) + k_{\rm dr} n_e}, \eqno(3)$$
where $k_{op}$ is the rate coefficient for the forward reaction in (2), $k_{po}$ is the rate coefficient for the reverse reaction, and $k_{\rm dr}$ is the rate coefficient for dissociative recombination of $\rm H_2Cl^+$.
Now using the principle of detailed balance to infer that $k_{po} = k_{op} \rm OPR_{\rm LTE}(H_2Cl^+)$, we may obtain the chloronium OPR as a weighted mean of
$\rm OPR_{LTE}(H_2Cl^+)$ (black curve in Figure 7) and $\rm OPR_0(H_2Cl^+)$ (blue curve in Figure 7)
$${\rm OPR(H_2Cl^+)} = x\,{\rm OPR_{LTE}(H_2Cl^+)} + (1 - x) {\rm OPR_{0}(H_2Cl^+)}, \eqno(4)$$
where $x=k_{\rm op} / [k_{\rm op} n({\rm H}) + k_{\rm dr} n_e]$ lies between zero and unity.  For 100~K gas that is primarily atomic with an electron abundance equal to the gas-phase elemental abundance of carbon ($1.6 \times 10^{-4}$), we find that $k_{\rm dr} n_e / n({\rm H}) \sim 5 \times 10^{-11}\, \rm cm^{3}\,s^{-1}$.  Thus, if thermalization proceeds at the Langevin rate with $k_{op} \sim 10^{-9}$, then $x$ in equation (4) is roughly $0.95$ and the expected chloronium OPR is close to $\rm OPR_{\rm LTE}$; in the opposite limit where $k_{op} \ll 5 \times 10^{-11}\,\rm cm^{3}\,s^{-1}$, we find that $x \ll 1$ and the OPR is close to $\rm OPR_{\rm 0}.$  Once again, future studies will be needed to estimate the value of $k_{op}$. 

Given our assumptions about the possible errors, all of the observed chloronium OPR values plotted in Figure 5 (lower panel) are consistent with 3, and most are consistent with 2.5.  In light of the foregoing discussion, and given a strong {\it prior} from astrochemical modeling that chloronium in diffuse clouds is primarily present in gas with a temperature of at least 30~K (for which $\rm OPR_{\rm LTE}(H_2Cl^+)$ is very close to 3), an OPR between 2.5 and 3 can be explained in any of the following scenarios: (1) chloronium is formed by the hopping mechanism; (2) the rate coefficient $k_{op} \simgt 1.5 \times 10^{-10}\,\rm cm^{3}\,s^{-1}$ (such that $x \simgt 0.75$); or (3) the observed chloronium lies mainly in gas with $T \simgt 110$~K (such that $\rm OPR_{LTE}(H_2) \ge 1.8$ and $\rm OPR_{LTE}(H_2Cl^+) \ge 2.5$).  Further work will be needed to determine which of these scenarios applies (or apply).

\acknowledgments

HIFI has been designed and built by a consortium of institutes and university departments from across
Europe, Canada and the United States under the leadership of SRON Netherlands Institute for Space
Research, Groningen, The Netherlands and with major contributions from Germany, France and the US.
Consortium members are: Canada: CSA, U.~Waterloo; France: CESR, LAB, LERMA, IRAM; Germany:
KOSMA, MPIfR, MPS; Ireland, NUI Maynooth; Italy: ASI, IFSI-INAF, Osservatorio Astrofisico di Arcetri-
INAF; Netherlands: SRON, TUD; Poland: CAMK, CBK; Spain: Observatorio Astron\'omico Nacional (IGN),
Centro de Astrobiolog\'ia (CSIC-INTA). Sweden: Chalmers University of Technology - MC2, RSS \& GARD;
Onsala Space Observatory; Swedish National Space Board, Stockholm University - Stockholm Observatory;
Switzerland: ETH Zurich, FHNW; USA: Caltech, JPL, NHSC.

Support for this work was provided by NASA through an award issued by JPL/Caltech.  JRG thanks MINECO for funding support under grants CSD2009-00038, AYA2009-07304 and AYA2012-32032.

{}
\clearpage

\begin{table*}
\footnotesize
\caption{List of observed sources}
\begin{tabular}{lccccccc}

\\
\hline
Source & R.A. & Dec & Distance$^a$ & $v_{\rm sys}$ $^a$ & OBSID & \multicolumn{2}{c}{Continuum $T_{\rm A}$(DSB)} \\
	   & J2000 & J2000 & kpc & km~s$^{-1}$ & --1342268000 &  (485 GHz) & (781 GHz) \\
\hline

G29.96--0.02 	   	& 18:46:03.80       & --02:39:22.0 & 5.26  &  98  & 615-617 &         & 1.17 K \\
                  &                   &              &       &      & \re{459-461} & 0.25 K 		 & \\
W49N 				& 19:10:13.20 		&  +09:06:12.0 & 11.11 &  \re{11}   & 618-620 &  & 4.80 K \\
                    &                   &              &       &      & 456-458 & 1.24 K		 & \\
W51  				& 19:23:43.90  		&  +14:30:30.5 & 5.1   &  \re{58}  & 621-623 &  & 5.40 K\\
                      &                   &              &       &      & \re{453-455} & 	1.44 K	 & \\
W3(OH)              & 02:27:03.82		&  +61:52:24.6 & 2.04  & \re{--44} & 625-627 &  & 1.88 K \\
                     &                   &              &       &      & \re{470-472} & 	0.49 K	 & \\
\hline
\tablenotetext{a}{References for distance \re{and $v_{\rm sys}$} estimates in the order listed:
Zhang et al. (2014); Zhang et al. (2013); Sato et al. (2010); Hachisuka et al. (2006).  All \re{distance estimates} are trigonometric estimates \re{based on observations of water masers.}}
\end{tabular} 
\end{table*}

\begin{table*}
\caption{Chloronium line frequencies and spontaneous radiative rates}
\begin{tabular}{lllcc}
\\
\hline
Species & Line  && Frequency$^a$ & $A_{ul}$ $^{a,b}$ \\
        &       && (GHz) & (s$^{-1}$) \\
\hline
$\rm o-H_2^{35}Cl^+$ &$2_{12}-1_{01}$ & $F=5/2-5/2$ &        781.60886 & 
$ 1.79\times 10^{-3}$ \\
        .....        &     .....      & $F=3/2-1/2$ &        781.61075 & 
$ 2.48\times 10^{-3}$ \\
        .....        &     .....      & $F=3/2-5/2$ &        781.62125 & 
$ 2.98\times 10^{-4}$ \\
        .....        &     .....      & $F=5/2-3/2$ &        781.62240 & 
$ 4.17\times 10^{-3}$ \\
        .....        &     .....      & $F=7/2-5/2$ &        781.62653 & 
$ 5.96\times 10^{-3}$ \\
        .....        &     .....      & $F=1/2-1/2$ &        781.62820 & 
$ 4.96\times 10^{-3}$ \\
        .....        &     .....      & $F=3/2-3/2$ &        781.63479 & 
$ 3.18\times 10^{-3}$ \\
        .....        &     .....      & $F=1/2-3/2$ &        781.65224 & 
$ 9.93\times 10^{-4}$ \\
\hline
$\rm o-H_2^{37}Cl^+$ &$2_{12}-1_{01}$ & $F=5/2-5/2$ &        780.03907 & 
$ 1.78\times 10^{-3}$ \\
        .....        &     .....      & $F=3/2-1/2$ &        780.04058 & 
$ 2.47\times 10^{-3}$ \\
        .....        &     .....      & $F=3/2-5/2$ &        780.04886 & 
$ 2.96\times 10^{-4}$ \\
        .....        &     .....      & $F=5/2-3/2$ &        780.04977 & 
$ 4.14\times 10^{-3}$ \\
        .....        &     .....      & $F=7/2-5/2$ &        780.05305 & 
$ 5.92\times 10^{-3}$ \\
        .....        &     .....      & $F=1/2-1/2$ &        780.05437 & 
$ 4.93\times 10^{-3}$ \\
        .....        &     .....      & $F=3/2-3/2$ &        780.05956 & 
$ 3.16\times 10^{-3}$ \\
        .....        &     .....      & $F=1/2-3/2$ &        780.07335 & 
$ 9.87\times 10^{-4}$ \\
\hline
$\rm p-H_2^{35}Cl^+$ &$1_{11}-0_{00}$ & $F=3/2-3/2$ &        485.41343 & 
$ 1.59\times 10^{-3}$ \\
        .....        &     .....      & $F=5/2-3/2$ &        485.41767 & 
$ 1.59\times 10^{-3}$ \\
        .....        &     .....      & $F=1/2-3/2$ &        485.42080 & 
$ 1.59\times 10^{-3}$ \\
\hline

\hline
\end{tabular}
\tablenotetext{a}{Based on laboratory spectroscopy of Araki (2001).  Downloaded from the CDMS database on 2015 Mar 25.}
\tablenotetext{b}{For an assumed dipole moment $\mu=1.89$~Debye (M\"uller 2008).  The spontaneous radiative rates are proportional to $\mu^2$.}
\end{table*}

\clearpage
\begin{table*}
\tiny
\caption{Chloronium column densities and abundances}
\begin{tabular}{llccccc}
\\
\hline
Source & $v_{\rm LSR}$  & $N(\rm{o-H_2^{35}Cl^+})$ & $N(\rm{H_2Cl^+})$$^a$  & $\Delta N(\rm{H_2Cl^+})$$^b$ & $N(\rm{H}$)$^c$ & $N(\rm{H_2Cl^+})/N(\rm{H})$  \\
& (km~s$^{-1}$) & ($10^{12}\,\rm{cm}^{-2}$) & ($10^{12}\,\rm{cm}^{-2}$) &  ($10^{12}\,\rm{cm}^{-2}$) & ($10^{21}\,\rm{cm}^{-2}$) & $\times \, 10^{9}$  \\
\hline
 G29.96-0.02 &       $[-10 ,  6]$ &   1.29 &   2.28 &   2.62 &   1.40 &   1.63     \\
    ...      &       $[  6 , 17]$ &   2.68 &   4.73 &   2.62 &   3.45 &   1.37     \\
    ...      &       $[ 17 , 28]$ &   0.99 &   1.74 &   2.62 &   1.42 &   1.22     \\
    ...      &       $[ 56 , 65]$ &   3.85 &   6.79 &   2.62 &   1.62 &   4.19     \\
    ...      &       $[ 65 , 73]$ &   4.33 &   7.63 &   2.62 &   2.32 &   3.29     \\
    ...      &       $[ 73 , 79]$ &   1.18 &   2.07 &   2.62 &   1.69 &   1.23     \\
    ...      &       $[ 79 , 88]$ &   1.49 &   2.63 &   2.62 &   2.58 &   1.02     \\
    ...      &   $[ 88 , 95]$$^d$ &   2.00 &   3.52 &   2.62 &   1.84 &   1.91$^d$ \\
    ...      &   $[ 95 ,113]$$^d$ &  11.45 &  20.19 &   2.62 &   6.02 &   3.35$^d$ \\
        W49N &   $[  1 , 10]$$^d$ &   7.16 &  12.63 &   0.64 &   4.44 &   2.84$^d$ \\
    ...      &   $[ 10 , 17]$$^d$ &  12.79 &  22.55 &   0.64 &   3.58 &   6.30$^d$ \\
    ...      &       $[ 17 , 25]$ &   4.55 &   8.03 &   0.64 &   2.34 &   3.43     \\
    ...      &       $[ 25 , 43]$ &  15.88 &  28.01 &   0.64 &   5.58 &   5.02     \\
    ...      &       $[ 43 , 51]$ &   3.35 &   5.91 &   0.64 &   1.67 &   3.54     \\
    ...      &       $[ 51 , 66]$ &  12.07 &  21.29 &   0.64 &   6.15 &   3.46     \\
    ...      &       $[ 66 , 80]$ &   3.68 &   6.48 &   0.64 &   2.89 &   2.24     \\
         W51 &       $[ 11 , 16]$ &   1.49 &   2.63 &   0.57 &   0.84 &   3.13     \\
    ...      &       $[ 16 , 21]$ &   0.45 &   0.79 &   0.57 &   0.88 &   0.90     \\
    ...      &       $[ 21 , 33]$ &   1.65 &   2.91 &   0.57 &   1.42 &   2.05     \\
    ...      &       $[ 33 , 42]$ &   1.02 &   1.79 &   0.57 &   0.84 &   2.13     \\
    ...      &   $[ 42 , 55]$$^d$ &   7.27 &  12.82 &   0.57 &   4.43 &   2.89$^d$ \\
    ...      &   $[ 55 , 62]$$^d$ &   3.29 &   5.80 &   0.57 &   3.25 &   1.79$^d$ \\
    ...      &       $[ 62 , 75]$       &  11.85       &  20.90       &   0.57    & $ >  1.54  $ & $ < 13.57$\\
        W3OH &   $[-51 ,-39]$$^d$ &   8.06 &  14.22 &   1.63 &   1.30 &  10.94$^d$ \\
    ...      &       $[-25 , -8]$ &   2.63 &   4.63 &   1.63 &   1.39 &   3.33     \\
    ...      &       $[ -8 ,  9]$ &   2.41 &   4.25 &   1.63 &   0.95 &   4.47     \\

\hline
\end{tabular}
\tablenotetext{a}{Total chloronium column density (both symmetry states and both isotopologues),
for an assumed $N(\rm{H_2^{35}Cl^+})/N(\rm{H_2^{37}Cl^+})$ ratio of 3.1 and an assumed OPR of 3}
\tablenotetext{b}{Amount by which $N(\rm{H_2Cl^+})$ is underestimated if a weak emission line of integrated intensity 250 mK $\rm km\,s^{-1}$ is present in the velocity interval}
\tablenotetext{c}{From Indriolo et al.\ (2015), based on HI 21 cm observations by B.\ Winkel et al.\ (2015, in preparation)}
\tablenotetext{d}{Velocity intervals close to the source systemic velocity, for which the $N(\rm{H_2Cl^+})$ estimates are unreliable} 

\end{table*}

\clearpage
\begin{table*}
\small
\caption{Chloronium column densities, isotopic ratios and ortho-to-para ratios toward W49N}
\begin{tabular}{lccccc}
\\
\hline
$v_{\rm LSR}$  & $N(\rm{o-H_2^{35}Cl^+})$ & $N(\rm{o-H_2^{37}Cl^+})$  &  $N(\rm{p-H_2^{35}Cl^+})$ & $N(\rm{H_2^{35}Cl^+})/N(\rm{H_2^{37}Cl^+})$ & OPR \\
(km~s$^{-1}$) & ($10^{12}\,\rm{cm}^{-2}$) & ($10^{12}\,\rm{cm}^{-2}$) &  ($10^{12}\,\rm{cm}^{-2}$) & Note (a) & Note (a)  \\
\hline
  $[  1 , 10]$$^b$ &   7.16 &   2.36 &   0.64 &  $ 3.04^{+ 0.21}_{-0.52}$ &  $11.16^{+ 0.75}_{-4.62}$\\
  $[ 10 , 17]$$^b$ &  12.79 &   3.62 &   2.69 &  $ 3.53^{+ 0.13}_{-0.42}$ &  $ 4.76^{+ 0.18}_{-0.69}$\\
      $[ 17 , 25]$ &   4.55 &   0.67 &   1.15 &  $ 6.81^{+ 0.72}_{-2.87}$ &  $ 3.98^{+ 0.42}_{-1.13}$\\
      $[ 25 , 43]$ &  15.88 &   4.06 &   4.55 &  $ 3.91^{+ 0.12}_{-0.42}$ &  $ 3.49^{+ 0.11}_{-0.32}$\\
      $[ 43 , 51]$ &   3.35 &   1.16 &   1.20 &  $ 2.89^{+ 0.42}_{-0.86}$ &  $ 2.80^{+ 0.40}_{-0.77}$\\
      $[ 51 , 66]$ &  12.07 &   4.06 &   3.79 &  $ 2.97^{+ 0.12}_{-0.32}$ &  $ 3.19^{+ 0.13}_{-0.34}$\\
      $[ 66 , 80]$ &   3.68 &   1.36 &   1.23 &  $ 2.70^{+ 0.36}_{-0.71}$ &  $ 2.99^{+ 0.39}_{-0.80}$\\
\hline
        $[17, 80]$ &  39.54 &  11.31 &  11.91 &  $ 3.50^{+ 0.21}_{-0.62}$ &  $ 3.32^{+ 0.20}_{-0.53}$\\

\hline
\end{tabular}
\tablenotetext{a}{Error bounds indicate the errors that could result from the presence of 
a weak interloper emission line within the velocity interval of integrated intensity 250 mK $\rm km\,s^{-1}$ (125 mK $\rm km\,s^{-1}$ at the $\rm {p-H_2^{35}Cl^+}$ frequency).}
\tablenotetext{b}{Velocity intervals close to the source systemic velocity, for which the $N(\rm{H_2Cl^+})$ estimates are unreliable} 
\end{table*}

\begin{figure*}
\centering
\includegraphics[width= 12 cm]{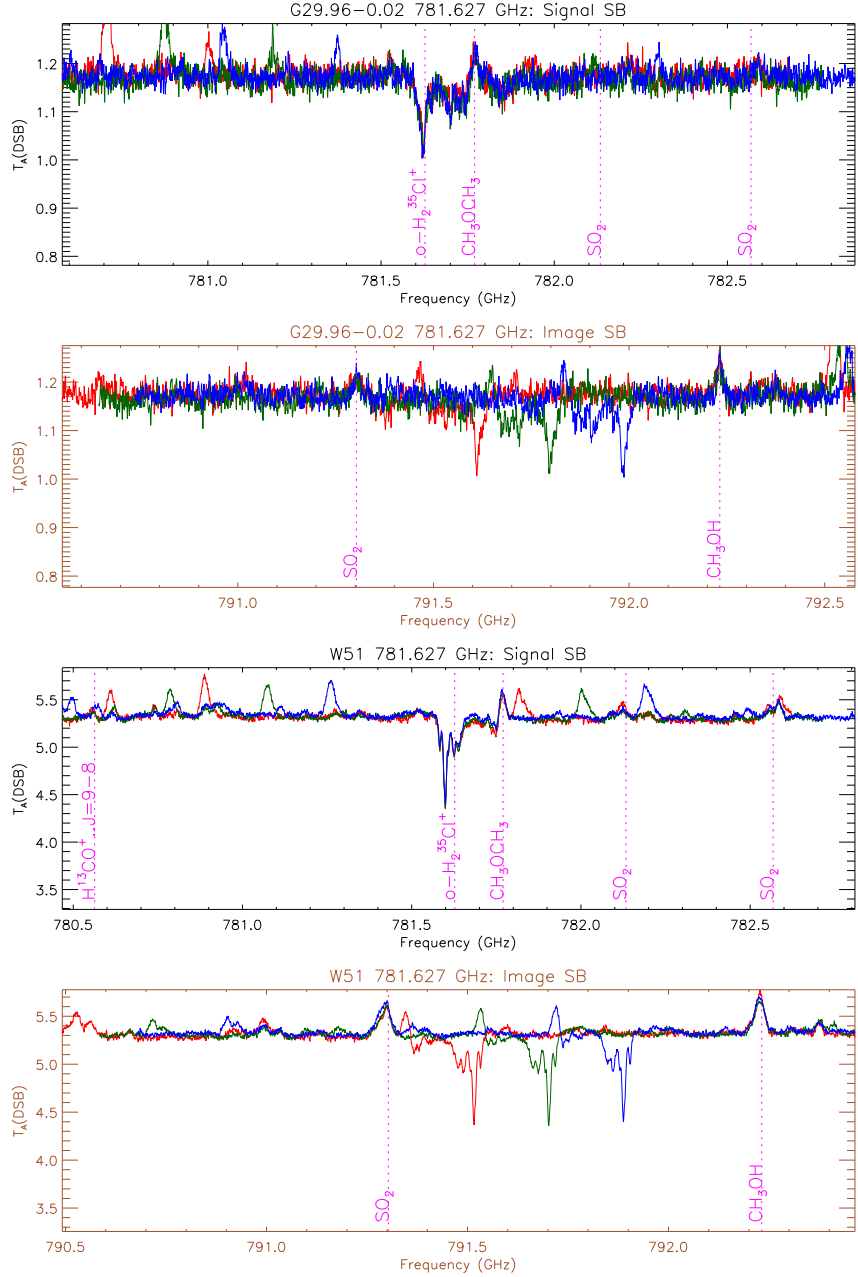}
\vskip 0.2 true in
\caption{Spectra for  $2_{12} - 1_{01}$ transition of ortho-$\rm H_2^{35}Cl^+$ obtained at three LO settings are shown separately in red, green and blue, with the frequency scale computed for either the signal or image sideband (alternate panels) in the rest frame of the source.  Magenta dashed lines indicate identified spectral lines in the relevant sideband.  Top two panels: G29.96 -- 0.02, for an assumed source velocity of +98 $\rm km\,s^{-1}$ relative to the LSR.  Bottom two panels: W51, for an assumed source velocity of +55 km~s$^{-1}$ relative to the LSR.}
\end{figure*}

\begin{figure*}
\centering
\includegraphics[width= 12 cm]{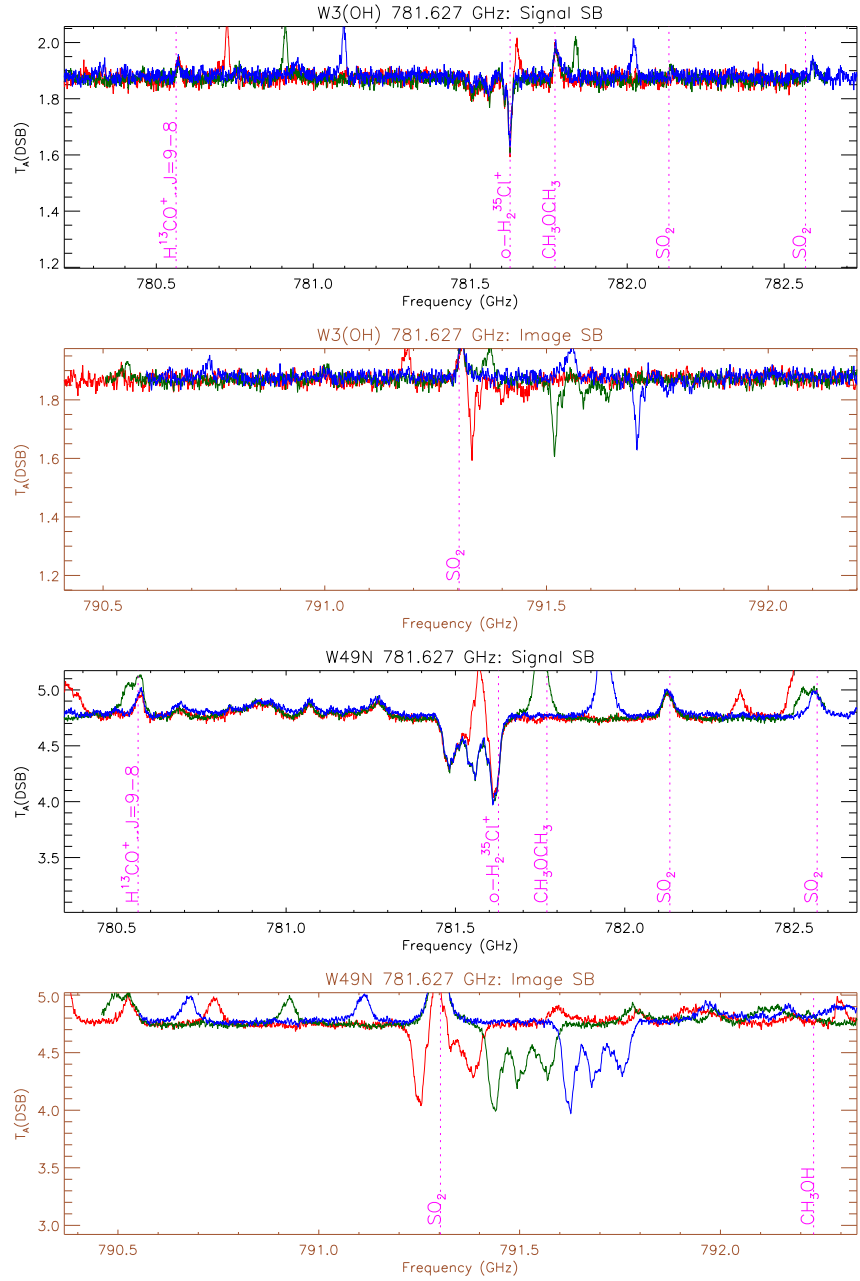}
\vskip 0.2 true in
\caption{Same as Figure 1, except for W3(OH) (top two panels, for an assumed source velocity of --45 km~s$^{-1}$ relative to the LSR) and W49N (bottom two panels, for an assumed source velocity of +8 km~s$^{-1}$ relative to the LSR).}
\end{figure*}

\begin{figure*}
\centering
\includegraphics[width= 12 cm]{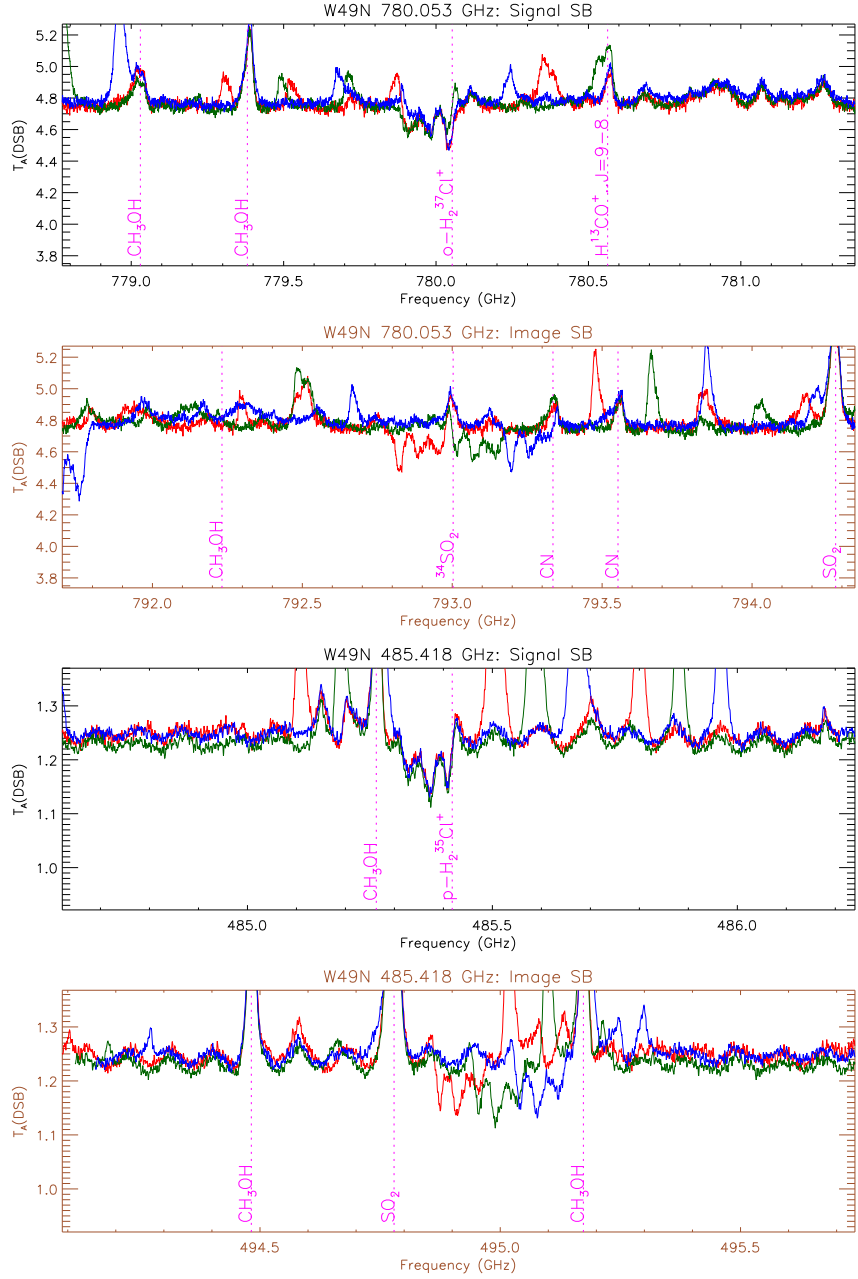}
\vskip 0.2 true in
\caption{Same as Figure 1, except for the ortho-$\rm H_2^{37}Cl^+$ $2_{12} - 1_{01}$ (top two panels) and para-$\rm H_2^{37}Cl^+$ $1_{11} - 0_{00}$ transitions (bottom two panels) observed toward W49N.}
\end{figure*}

\begin{figure*}
\centering
\includegraphics[width= 13 cm]{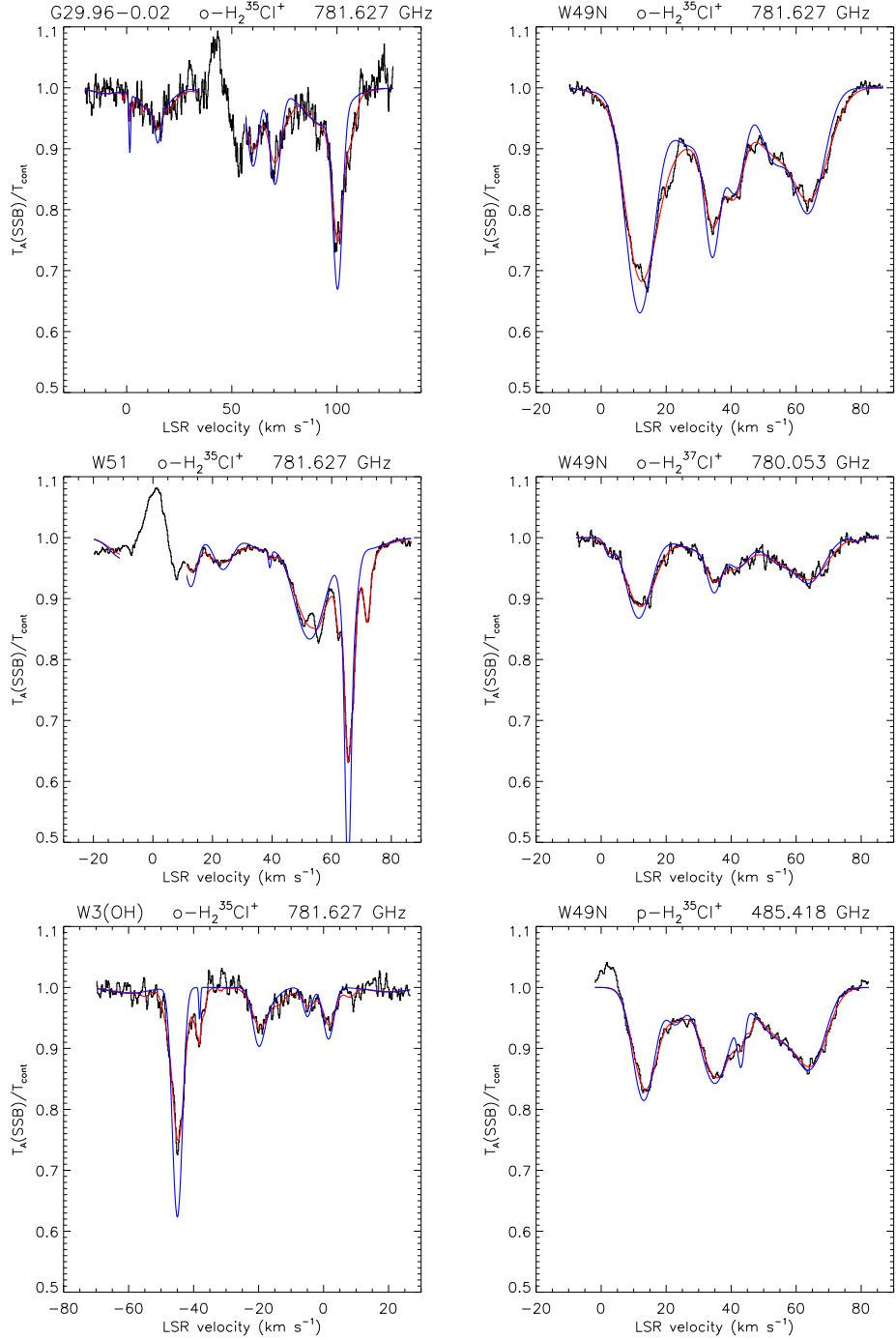}
\vskip 0.2 true in
\caption{Transmission fractions for the  ortho-$\rm H_2^{35}Cl^+$ $2_{12} - 1_{01}$ transition observed toward G29.96 -- 0.02, W51, W3(OH) and W49N; and for the  ortho-$\rm H_2^{37}Cl^+$ $2_{12} - 1_{01}$
and \re{para-$\rm H_2^{35}Cl^+$} $1_{11} - 0_{00}$ transitions observed toward W49N.  Black histogram: observed data.  Red curve: fit to data.  Blue curve: hyperfine-deconvolved fit.  The emission lines in the G29.96 -- 0.02 and W51 panels are due to dimethyl ether.}
\end{figure*}

\begin{figure*}
\centering
\includegraphics[width= 12 cm]{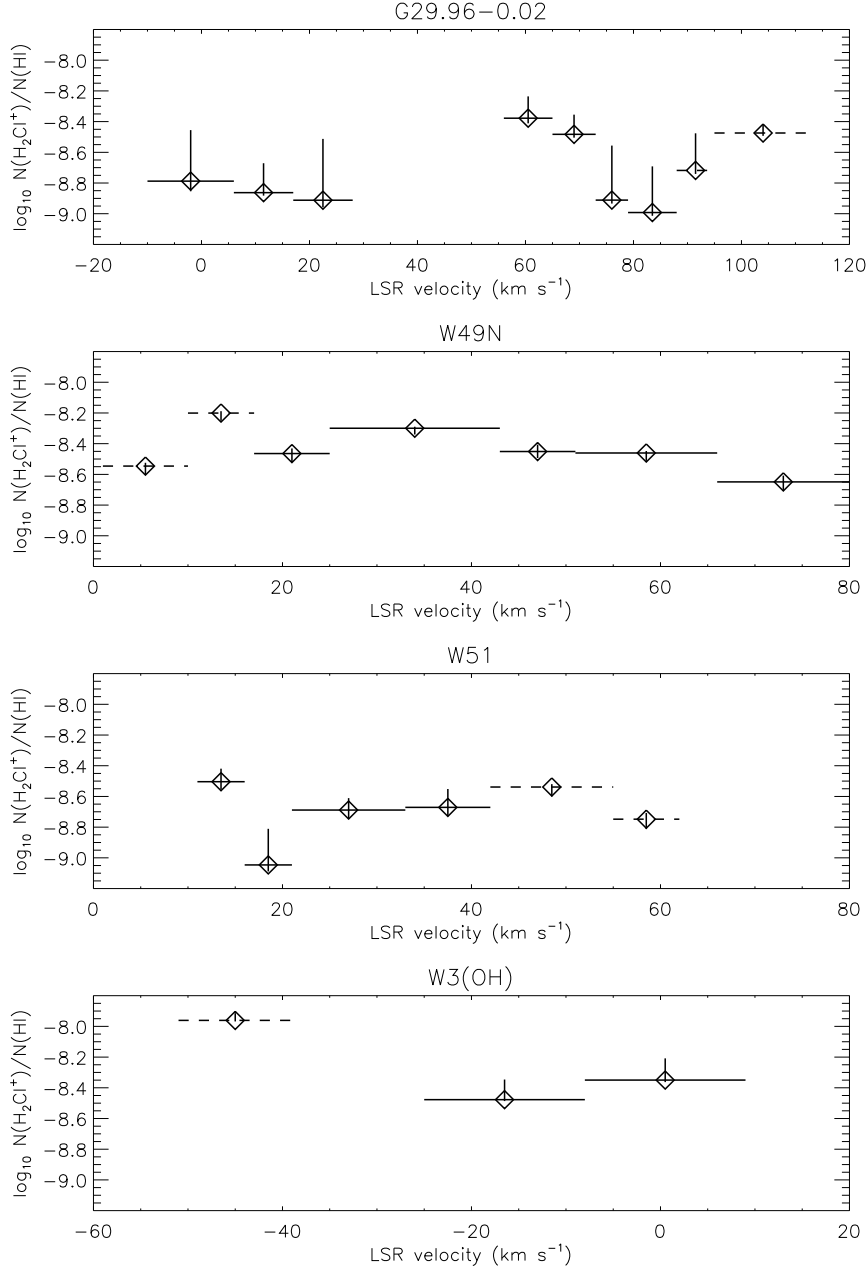}
\vskip 0.2 true in
\caption{Inferred $N({\rm H_2Cl^+})/N({\rm H})$ ratios for the velocity intervals adopted by Indriolo et al.\ (2015).  Dashed horizontal bars indicate velocity intervals close the source systemic velocity, where the column density determinations are unreliable.  Vertical error bars indicate uncertainties, and include the possibility that weak emission lines (with a width of 5 km~s$^{-1}$ an assumed maximum antenna temperature of 50 mK) result in an underestimate of the inferred $\rm H_2Cl^+$ optical depth.} 
\end{figure*}

\begin{figure*}
\centering
\includegraphics[width= 11 cm]{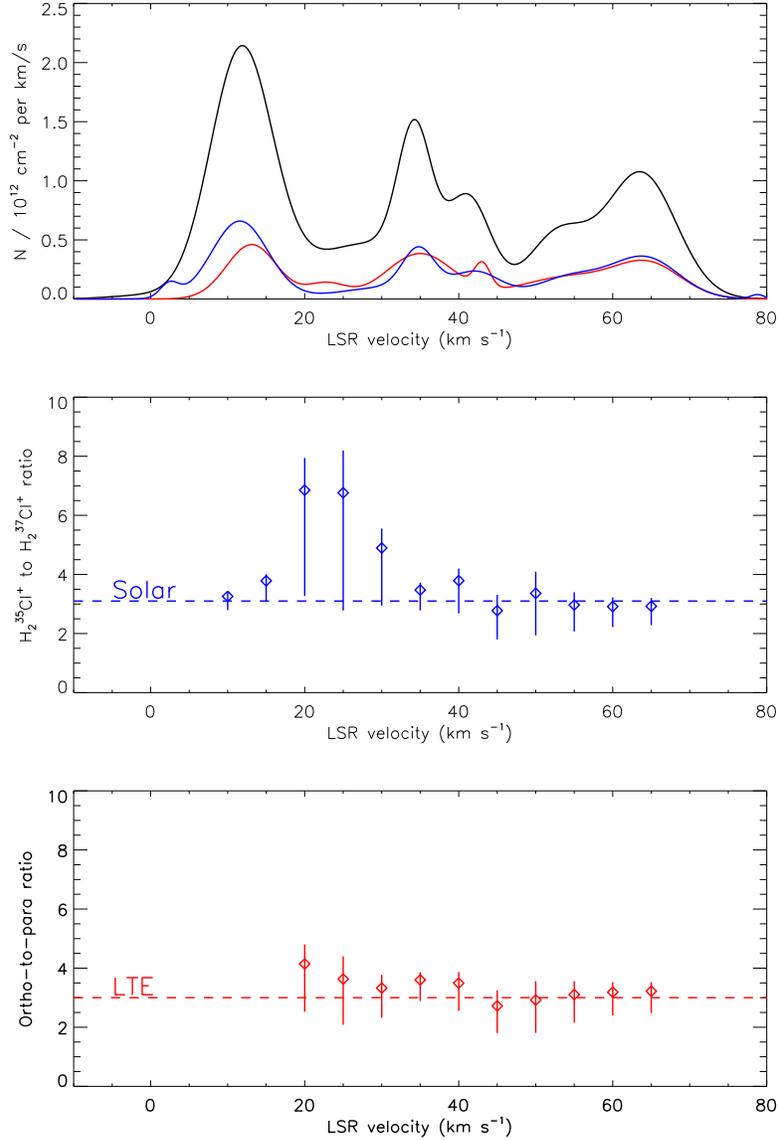}
\vskip 0.2 true in
\caption{Top panel: derived column densities for ortho-$\rm H_2^{35}Cl^+$ (black), ortho-$\rm H_2^{37}Cl^+$ (blue), and para-$\rm H_2^{35}Cl^+$ (red) along the sight-line to W49N.   Values close the systematic velocity of the source (8 km~s$^{-1}$) are unreliable.  Middle panel: inferred ortho-$\rm H_2^{35}Cl^+$ / ortho-$\rm H_2^{37}Cl^+$ ratio in velocity bins of width 5 km~s$^{-1}$.  Bottom panel: inferred ortho-to-para ratio for $\rm H_2^{35}Cl^+$ in velocity bins of width 5 km~s$^{-1}$. Vertical error bars in the bottom two \re{panels} indicate uncertainties, and include the possibility of weak emission lines (with an assumed width of 5 km~s$^{-1}$ and maximum antenna temperature of 50 mK near 781 GHz or 25 mK near 485 GHz) resulting in an underestimate of the inferred optical depths.}   
\end{figure*}

\begin{figure*}
\centering
\includegraphics[width= 16 cm]{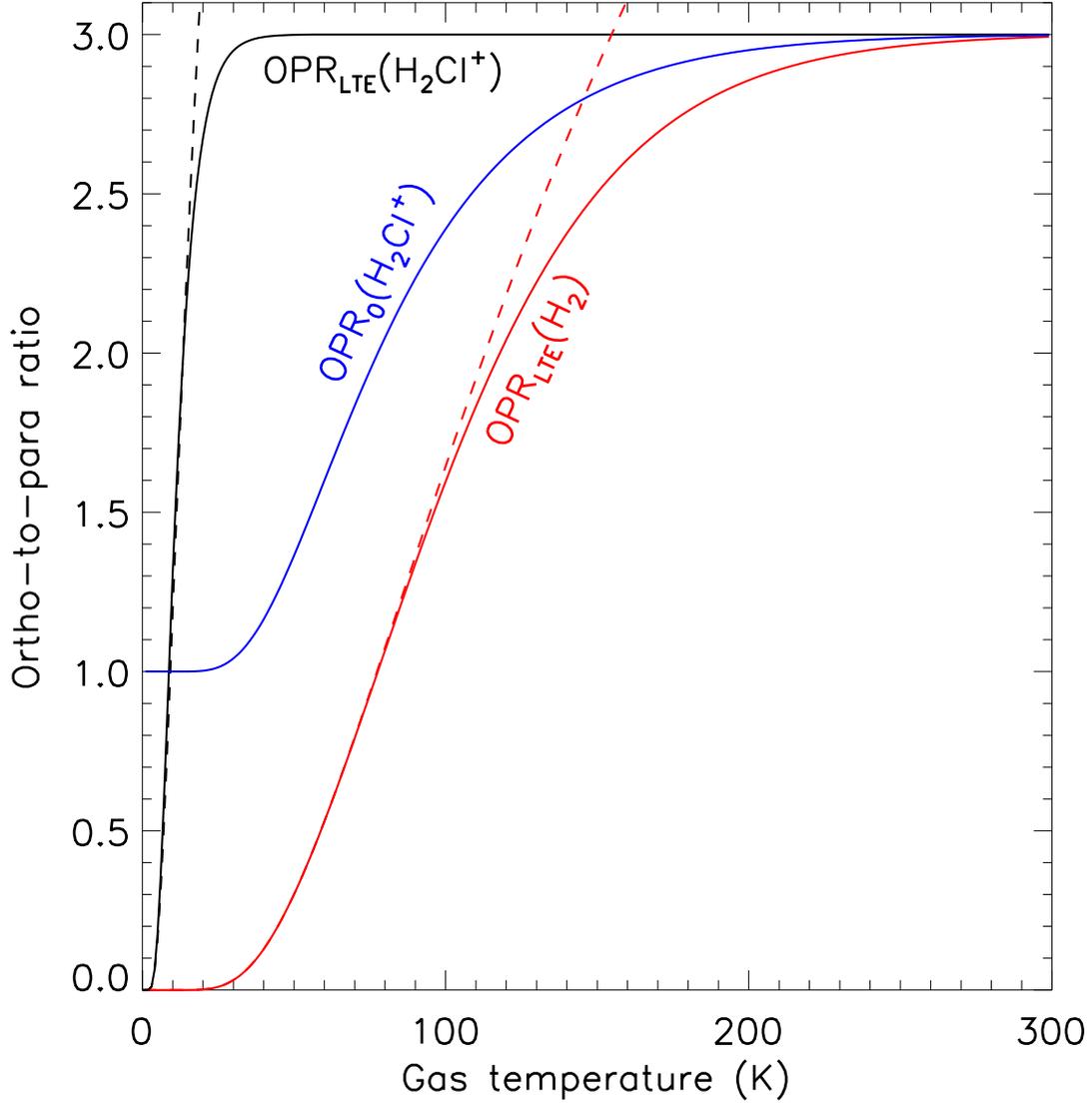}
\vskip 0.2 true in
\caption{Ortho-to-para ratios (OPR) as a function of gas temperature.  Solid black curve: 
$\rm H_2Cl^+$ OPR in LTE. Dashed black curve: $9\, \exp(-20.2 {\rm K}/T)$.  Solid red curve: $\rm H_2$ OPR in LTE. Dashed red curve: $9\, \exp(-170 {\rm K}/T)$.  Blue curve: initial OPR of $\rm H_2Cl^+$ formed by reaction of HCl$^+$ with H$_2$, for the scrambling mechanism (see text) and with the H$_2$ OPR in LTE. }
\end{figure*}

\end{document}